\documentclass[letterpaper, 10 pt, conference]{ieeeconf}  % Comment this line out if you need a4paper

\IEEEoverridecommandlockouts                              % This command is only needed if 
\usepackage{amsmath} % assumes amsmath package installed
\usepackage{amssymb}  % assumes amsmath package installed
\usepackage{xcolor} 
\usepackage{graphicx}
\usepackage{booktabs}
\usepackage{bm}

\title{\LARGE \bf
Combining Deep Reinforcement Learning with a Jerk-Bounded Trajectory Generator for Kinematically Constrained Motion Planning*
}

\author{Seyed Adel Alizadeh Kolagar, Mehdi Heydari Shahna, and Jouni Mattila% <-this % stops a space
\thanks{*Funding for this research was provided by the Business Finland partnership project ``Future All-Electric Rough Terrain Autonomous Mobile Manipulators'' (Grant No. 2334/31/2022).}% <-this % stops a space
\thanks{All authors are with the Faculty of Engineering and Natural Sciences, Tampere University, Finland
        {\{seyedadel.alizadehkolagar; mehdi.heydarishahna; jouni.mattila\}@tuni.fi}}%
}

\begin{document}

\maketitle
\thispagestyle{empty}
\pagestyle{empty}

%%%%%%%%%%%%%%%%%%%%%%%%%%%%%%%%%%%%%%%%%%%%%%%%%%%%%%%%%%%%%%%%%%%%%%%%%%%%%%%%
\begin{abstract}

Deep reinforcement learning (DRL) is emerging as a promising method for adaptive robotic motion and complex task automation, effectively addressing the limitations of traditional control methods. 
However, ensuring safety throughout both the learning process and policy deployment remains a key challenge due to the risky exploration inherent in DRL, as well as the discrete nature of actions taken at intervals. These discontinuities, despite being part of a continuous action space, can lead to abrupt changes between successive actions, causing instability and unsafe intermediate states.
To address these challenges, this paper proposes an integrated framework that combines DRL with a jerk-bounded trajectory generator (JBTG) and a robust low-level control strategy, significantly enhancing the safety, stability, and reliability of robotic manipulators. The low-level controller ensures the precise execution of DRL-generated commands, while the JBTG refines these motions to produce smooth, continuous trajectories that prevent abrupt or unsafe actions.
The framework also includes pre-calculated safe velocity zones for smooth braking, preventing joint limit violations and ensuring compliance with kinematic constraints.
This approach not only guarantees the robustness and safety of the robotic system but also optimizes motion control, making it suitable for practical applications. The effectiveness of the proposed framework is demonstrated through its application to a highly complex heavy-duty manipulator.


\end{abstract}

%%%%%%%%%%%%%%%%%%%%%%%%%%%%%%%%%%%%%%%%%%%%%%%%%%%%%%%%%%%%%%%%%%%%%%%%%%%%%%%%
\section{INTRODUCTION}

Deep reinforcement learning (DRL), recognized for its success in Go and video games \cite{silver2017mastering, mnih2015human}, is becoming a promising approach for adaptive robotic motions and complex task automation \cite{gu2017deep}. It addresses the limitations of traditional methods, such as model inaccuracies and high stochasticity, but often involves risky random exploration. Consequently, DRL is usually tested in simulators \cite{brockman2016openai} or controlled environments to ensure safety \cite{levine2018learning, akkaya2019solving}. 

A key challenge in robot learning is maintaining safety during both the training process and policy deployment. Provably safe reinforcement learning (DRL) tackles this by integrating system knowledge to ensure that the agent explores only safe actions and states \cite{garcia2015comprehensive}. State-wise approaches, such as State-Wise Constrained Markov Decision Processes (SCMDPs) \cite{berkenkamp2017safe}, provide formal safety guarantees \cite{zhao2023state}. One effective method is action masking, which proactively filters out unsafe actions, preserving the policy's integrity and avoiding the issues associated with action replacement or projection, where actions are modified post-selection \cite{krasowski2023provably}.

Recent advancements in DRL enable robots to autonomously learn and plan optimal trajectories in unstructured environments, with model-free DRL algorithms being particularly effective for motion planning \cite{xie2019deep}.
Higher-level action space choices and using lower-level stabilizing controllers typically result in better trajectory optimization and policy performance, often outperforming end-to-end policies \cite{kaufmann2022benchmark}.
Studies on action space suggest using joint velocities as actions, which requires tracking by a separate low-level controller \cite{aljalbout2024role}. However, in DRL algorithms, the generated actions are discrete at intervals (despite continuous action space), making the controller's performance dependent on the changes between successive actions and the time intervals. This can lead to instability due to discontinuities and abrupt changes. Increasing the DRL step time can help the controller with stabilization but may reduce the DRL performance by lowering the observation frequency, potentially leading to unsafe intermediate states.
A commonly used method for smoothing sudden trajectory changes is the trapezoidal velocity profile, valued for its optimality under constraints. However, it introduces infinite jerk, which complicates tracking and increases motion time \cite{macfarlane2003jerk}.
To address these issues, we propose incorporating a jerk-bounded profile into the generated actions. For that purpose, we developed a new jerk-bounded trajectory generator (JBTG) for the DRL framework inspired by \cite{macfarlane2003jerk} that will be used to smooth the sudden changes between actions generated by the DRL to satisfy kinematic constraints in each DRL step. 
Jerk-bounded profiles are preferred as they minimize vibrations, prevent the excitation of natural frequencies (particularly crucial for heavy-duty robots in which even low frequencies of motion can produce significant instantaneous accelerations \cite{vihonen2017joint}), and enhance tracking accuracy \cite{fang2020approach}. Polynomial interpolation is commonly used for trajectory planning with waypoints, with fifth-order polynomials being the lowest degree that satisfies third-order motion constraints \cite{macfarlane2003jerk}, although higher degrees may cause oscillations and instability \cite{liu2013time}.

\begin{figure*}[htpb] % Adjust the value as needed
    \centering
     \hspace{-1.15cm}\includegraphics[width=0.9\textwidth, height=3cm]{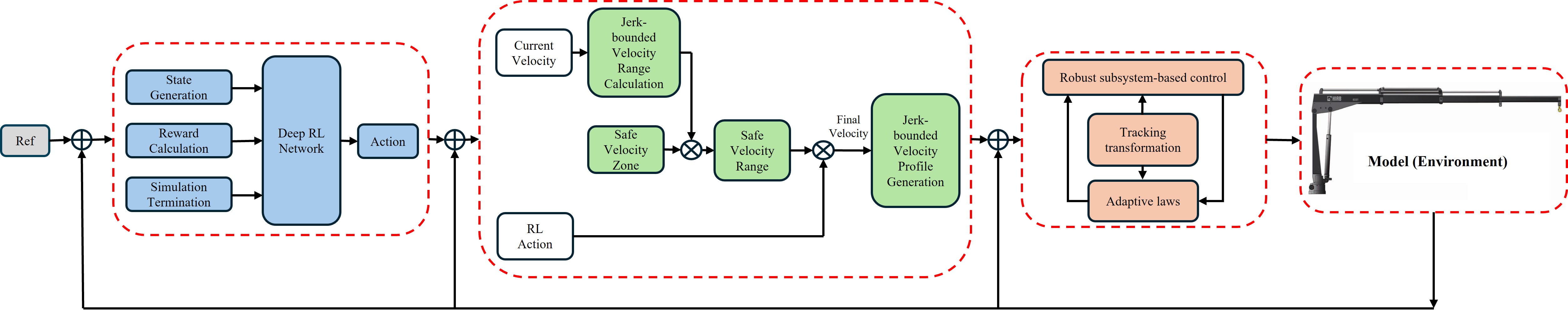}
    \caption{Architecture of the jerk-bounded DRL-based control of a HIAB manipulator}
    \label{Architecture}
\end{figure*}

To prevent the robot joints from exceeding position limits, we pre-calculate a safe velocity zone for each joint offline based on their specifications. The final safe motion combines the allowable velocity range from the JBTG with the velocities within this safe zone, ensuring that position limits are not violated and enabling the implementation of alternative safe behaviors in subsequent steps.

Following the shifting trend from end-to-end learning to methods that integrate established control principles for the implementation of the DRL-generated commands, many studies have adopted low-level control using classical controllers such as PD and PID \cite{lawrence2022deep, kumar2021joint, rudin2022learning}. However, these classical controllers often lack guaranteed robustness and stability, which can negatively impact system performance, especially in high-order dynamical systems \cite{carlucho2020adaptive, yang2021hierarchical}. Therefore, to ensure that the robot manipulator robustly tracks the trajectory generated by the JBTG, this paper employs a modern and sophisticated control strategy for low-level control. This strategy compensates for uncertainties and external disturbances at the joint level and ensures strong convergence toward the trajectory generated by the JBTG.

The proposed approach of integrating a DRL agent with a JBTG as motion planner and using a robust low-level control offers multiple benefits:

\begin{itemize}

\item Firstly, the controller now accesses a continuous velocity profile instead of discrete commands, making it easier to track and ensuring stability.

\item Secondly, the JBTG accounts for the kinematic limitations of the manipulator joints, ensuring that the generated trajectories are feasible for the manipulator to follow. In addition to being jerk-bounded, the trajectories are also jerk-continuous, ensuring smoother motion.

\item Lastly, by considering joint limits, we gain insight into future possible actions, which can be utilized for safety calculations.

\end{itemize}

The paper is structured as follows: Section 2 covers background concepts, Section 3 details the mathematical framework, Section 4 presents the simulations and results, and Section 5 summarizes key findings and future directions.

%%%%%%%%%%%%%%%%%%%%%%%%%%%%%%%%%%%%%%%%%%%%%%%%%%%%%%%%%%%%
\section{PRELIMINARIES AND BACKGROUND}

The overall procedure for controlling the robot using an integrated approach that combines DRL as the high-level controller and a robust low-level controller is illustrated in Fig. \ref{Architecture}. Initially, the DRL agent receives sensory data to compute the states. Based on these observations, the agent generates the next action within the interval of [-1, 1], which will then be mapped to the range between minimum and maximum velocities. These velocities are calculated to ensure safety and compliance with all kinematic constraints associated with using a JBTG. This framework determines the maximum and minimum allowable velocities for the next step while adhering to acceleration and jerk limits. Additionally, a safety framework adjusts the calculated velocity range to prevent violations of position constraints that could cause damage. The resulting safe velocity is then fed into a low-level controller for execution. Each of these frameworks will be further explained in the following sections.

\subsection{DRL Framework}
The goal of our DRL agent is to generate motion plans for a manipulator with n degrees of freedom (DoF), while guaranteeing kinematically safe joint-level commands.
Among the model-free DRL algorithms that have demonstrated promising results in motion planning \cite{xie2019deep}, the Soft Actor-Critic (SAC) algorithm distinguishes itself for its superior performance in high-dimensional problem-solving \cite{prianto2021deep}.

In this project, which focuses on trajectory planning for a reaching task of a heavy-duty robotic manipulator, we define the state, $s_t$, as follows:
\begin{equation}
\small
\begin{aligned}
\label{equation: 4a}
s_t=\left\langle f_t, p_t, \dot{p}_t, x_t, \dot{x}_t, p, \delta p_t,\right\rangle
\end{aligned}
\end{equation}
here, $f_t$ represents the joint forces, $p_t$ the joint positions (including joint values and boom angles), and $\dot{p}_t$ the corresponding velocities. $x_t$ and $\dot{x}_t$ denote the tip's position and velocity, respectively, while $p$ specifies the target location in Cartesian space. Finally, $\delta p_t$ is the distance between the end-effector and the target.
We chose joint velocities for the action space, $a_t$, due to their superior performance and ability to avoid kinematic singularities \cite{aljalbout2024role}. The agent outputs normalized values, which are then scaled to fit within a safe velocity range for execution.
\begin{equation}
\small
\begin{aligned}
\label{equation: 7a}
a_t=<v_t>
\end{aligned}
\end{equation}
where $v_t$ represents DRL-actuated joint velocities. 
The reward function consists of four terms. The first term penalizes the magnitude of the generated forces. The second term imposes a penalty based on the distance between the tip and the target position. The third term penalizes the magnitude of the tip velocity. Finally, the agent receives a positive reward for successfully reaching the target under the desired conditions.
To facilitate more efficient learning for the DRL agent, we employ curriculum learning (CL) \cite{narvekar2020curriculum}, allowing it to progressively acquire an optimized policy. In this approach, the robot is initially tasked with reaching targets at shorter distances. As the agent improves, the distance between the initial and target positions is gradually increased.

\subsection{Kinematic Constraints and Safety Considerations}
\label{safety}

To map actions to a safe velocity range, we must ensure both the current step and subsequent steps remain kinematically safe. This means the robot's joint motion must comply with the kinematic constraints on position, velocity, acceleration, and jerk \eqref{constraint1}.
\begin{equation}
\small
\begin{aligned}
\label{constraint1}
&p_{\text{min}} \leq p \leq p_{\text{max}}, \hspace{0.2cm} v_{\text{min}} \leq v \leq v_{\text{max}}\\
&a_{\text{min}} \leq a \leq a_{\text{max}}, \hspace{0.2cm} j_{\text{min}} \leq j \leq j_{\text{max}} 
\end{aligned}
\end{equation}

To ensure adherence to these conditions, we utilize a jerk-bounded velocity profile between DRL actions during each DRL time step. This approach guarantees that the system maintains the desired kinematic safety throughout its operation.
Unlike common JBTGs based on initial and final positions, our approach uses initial and final velocities and incorporates the time step of our DRL framework. We adapt the JBTG developed by \cite{macfarlane2003jerk} to our problem, ensuring safety with joint position limits and kinematic constraints.
Our DRL method requires low computational overhead for real-time execution. The method of using linked quintic polynomials \eqref{polynomial} ensures $C^2$ continuity and efficient, trackable motion. This approach follows a near-optimal template similar to that of the Linear Segment with Parabolic Blend (LSPB), respecting kinematic limits and avoiding oscillations. Using sine-shaped acceleration ramps complies with jerk limits, ensuring smooth and efficient trajectories.
\begin{equation}
\small
\begin{aligned}
\label{polynomial}
p(t)=b_5t^5+b_4t^4+b_3t^3+b_2t^2+b_1t+b_0
\end{aligned}
\end{equation}
The coefficients of \eqref{polynomial} must be defined to satisfy the required conditions in each quintic segment, accommodating non-zero initial and final velocities crucial for our DRL framework, where the action represents the reference velocity tracked by a controller. The trajectory generator computes up to eight quintic control points, ensuring adherence to speed, acceleration, and jerk limits, thereby making it suitable for real-time robotic applications and integrating safety and efficiency for modern robotic systems and DRL frameworks.

To achieve kinematic safety, we input the joint's minimum and maximum velocities into the JBTG and assess whether the desired velocities can be attained within a single step. If achievable, these values are designated as the allowable maximum and minimum velocities for that state. If not, the final reachable velocity is taken as the allowable velocity. This approach meets all constraints except for the position limits. 
This condition requires verifying not only the safety of the immediate next action but also preventing unsafe states in subsequent steps. Specifically, the robot must not approach its position limits at a velocity greater than zero in the direction of those limits. Following the methodology adopted by \cite{kiemel2021learning, kiemel2022learning}, a braking trajectory is computed for the generated velocity at each step. This involves a deceleration phase at the highest permissible deceleration value, ensuring zero velocity at zero acceleration and jerk. The computed trajectory and times determine the maximum allowable acceleration at that point, and the duration of the braking trajectory affects the computational effort.

To reduce the computational load, this paper proposes performing safety calculations offline. Since the joint kinematic constraints are known beforehand, calculations related to joint position limits are completed before training. At various distances from these limits, we pre-determine the maximum safe velocity, ensuring the joint can brake in time (considering acceleration and jerk limits) without exceeding position limits. As a result, the allowable velocity at any point is pre-calculated, ensuring compliance with position and kinematic constraints.

\subsection{Tracking Formulation}
After decomposing the manipulator into subsystems of joints, we can introduce position and velocity states ($\chi_1$ and $\chi_2$) for each joint as follows:
\begin{equation}
\small
\begin{aligned}
\label{300}
& \dot{\chi}_1(t)=\chi_2(t)+f_1 + d_1 \\
& \dot{\chi}_2(t)=A u(t)+f_2+ d_2
\end{aligned}
\end{equation}
where $f_1$ and $f_2$ represent modeling uncertainties. $d_1$ and $d_2$ signify load disturbances, and u is the control input signal. Now, we can define the tracking error as:
\begin{equation}
\small
\begin{aligned}
\label{302}
e_1(t) =& \chi_1(t) - x_r (t)\\
e_2 (t) =&  \chi_2(t) - \dot{x}_r (t)
\end{aligned}
\end{equation}
where $x_r$ is the jerk-bounded reference trajectory for each joint generated by DRL. By considering \eqref{300} and \eqref{302}, we have:
\begin{equation}
\small
\begin{aligned}
\label{303}
& \dot{e}_1(t)=e_2(t)+f_1 + d_1 \\
& \dot{e}_2(t)=A u(t)+f_2+ d_2  - \ddot{x}_r (t)
\end{aligned}
\end{equation}

%%%%%%%%%%%%%%%%%%%%%%%%%%%%%%%%%%%%%%%%%%%%%%%%%%%%%%%%%%%%
\section{Calculations and Mathematics}

\subsection{Jerk-Bounded Trajectory Generator}
\subsubsection{Terminology}

To precisely describe the trajectory segments, it is important to understand the following terms. A ramp refers to a change in any given parameter, while a cruise denotes a state where the parameter remains constant. A pulse is defined as a sequence with a ramp-up followed by a ramp-down. Lastly, a sustained pulse consists of a ramp-up, a cruise phase, and a subsequent ramp-down. 

\subsubsection{Sine-Shaped Trajectory Generation}
To generate a trajectory between two waypoints $(v_1,t_1)$ and $(v_2,t_2)$, we first explain a situation where $v_2=v_{\text{max}}$, and the required time for reaching $v_2$ from $v_1$ is less than the available time, giving it enough time to reach the maximum velocity. 

The initial and final velocities are denoted as \(v_1\) and \(v_2\), respectively, with \(t_1\) set to zero and \(t_2\) equal to \(dt\). To achieve the final velocity, four quintics are required. The first quintic ramps up to the maximum acceleration, resulting in a velocity of \(V_a\). The second is an acceleration cruise segment, culminating in a final velocity of \(V_b\). Following this, the acceleration ramp-down quintic brings the velocity down to \(v_2\) at zero acceleration. Lastly, the final quintic involves a velocity cruise, during which the robot joint maintains this velocity for the remaining time.
The required time for the ramp-up to $a_{\max }$ concerning acceleration and jerk limits can be calculated using \eqref{dtmax}. 
\begin{equation}
\small
\begin{aligned}
\label{dtmax}
d t_{\max }=\frac{\pi a_{\max }}{2 j_{\max }}
\end{aligned}
\end{equation}
The change in speed during acceleration ramp-up and ramp-down is the same and can be calculated using \eqref{v remaining}.
\begin{equation}
\small
\begin{aligned}
\label{v remaining}
dV_{\mathrm{ru}}=d V_{\mathrm{rd}}=\frac{a_{\mathrm{max}} d t_{\max }}{2}
\end{aligned}
\end{equation}
Using this equation, we can determine when the acceleration ramp-down begins.
The Modified Sustained Acceleration Pulse (MSAP) algorithm (see Algorithm 1) determines the four quintic control points for the speed ramp.
In some conditions, the MSAP algorithm cannot be used due to the limitations caused by time or desired speed changes. 
This happens when the time $dt$ is less than the time required for an acceleration pulse of $a_{\text{max}}$, which is $2\ \times$ $dt_{\text{max}}$.
In addition to that, if the desired change in speed, $dV=v_2-v_1$, is less than $dV_{\text{min}}$ \eqref{dvmin}, which is the speed change caused by an acceleration pulse of $a_{\text{max}}$, MSAP cannot be used. 
\begin{equation}
\small
\begin{aligned}
\label{dvmin}
d V_{min}=dt_{max}a_{max}
\end{aligned}
\end{equation}
\vspace{-0.5cm}

\begin{table}[h!]
  \centering
  \renewcommand{\arraystretch}{0.5} % Reset row height
  \colorbox{white!10}{\fbox{%
    \begin{tabular}{p{0.95\linewidth}} % 1 column, full width
      \multicolumn{1}{c}{\textbf{Algorithm 1.} Modified Sustained Acceleration Pulse (MSAP)} \\
      \\
      \textbf{Input}: $v_1, v_2, a_{max}, j_{max},$ and $dt$. 
      \\
      \\
      \textbf{Output}: Position, Velocity, Acceleration, and Jerk Profiles. 
      \\
      \begin{enumerate}
        \item Calculate speeds $V_a$ and $V_b$:
        $$
        V_a = v_1 + \frac{a_{\max} d t_{\max}}{2}, \quad V_b = v_2 - \frac{a_{\max} d t_{\max}}{2}.
        $$
        \item Calculate the necessary time for each quintic:
        \begin{enumerate}
          \item ramp-up time:
          $$
          t_1 = d t_{\max}.
          $$
          \item acceleration cruise time:
          $$
          t_2 = \frac{V_b - V_a}{a_{\max}}.
          $$
          \item ramp-down time:
          $$
          t_3 = d t_{\max}.
          $$
        \end{enumerate} 
        \item If $t_1 + t_2 + t_3 - d t > 0:$ 
        
        Reduce $v_2$ based on $d t$ and recalculate $t_2, t_3$, and $V_b$.
        \item Compute distances traveled during the quintics:
        \begin{enumerate}
          \item ramp-up distance to $a_{\max}$:
          $$
          D_1 = a_{\max} d t_{\max}^2 \left( \frac{1}{4} - \frac{1}{\pi^2} \right) + v_1 d t_{\max}.
          $$
          \item distance during acceleration cruise:
          $$
          D_2 = \frac{V_b^2 - V_a^2}{2 a_{\max}}.
          $$
          \item ramp-down distance from $a_{\max}$:
          $$
          D_3 = a_{\max} d t_{\max}^2 \left( \frac{1}{4} + \frac{1}{\pi^2} \right) + V_b d t_{\max}.
          $$
        \end{enumerate}
        \item Specify the time, position, velocity, acceleration, and jerk associated with the four quintic control points that constitute the segment of motion with nonzero acceleration.
        \vspace{-0.3cm}
      \end{enumerate}
    \end{tabular}%
  }}
\end{table}

In both scenarios, achieving the maximum allowable acceleration, $a_{\text{max}}$, without exceeding a defined trajectory limit is not feasible. Instead, the motion will be determined using an acceleration pulse with a reduced acceleration, \(a_{\text{peak}}\), which may be followed by a speed cruise.

If $dt<2\ \times$ $dt_{\text{max}}$, $dV>dV_{\text{min}}$, then only the time is limiting. In such a situation, we will have an acceleration pulse of $a_{\text{peak}}$ with the whole duration of $dt$. Using \eqref{apeak}, $a_{\text{peak}}$ can be calculated based on $dt_{\text{peak}}=dt/2$.
\begin{equation}
\small
\begin{aligned}
\label{apeak}
a_{\text {peak}}=\frac{2j_{\max }dt_{\text{peak}}}{\pi}
\end{aligned}
\end{equation}

If $dt>2\ \times$ $dt_{\text{max}}$, $dV<dV_{\text{min}}$, then the limiting item is the desired change in speed. The required time for reaching to $a_{\text{peak}}$ can be calculated using \eqref{apeak time}.
\begin{equation}
\small
\begin{aligned}
\label{apeak time}
dt_{\text{peak}}=\sqrt{\frac{\pi\left(v_2-v_1\right)}{2 j_{\max }}}
\end{aligned}
\end{equation}

The calculated $dt$ would then be used to calculate $a_{\text{peak}}$ with the help of \eqref{apeak}. In both cases, the first two quintics (three control points) are calculated for acceleration ramp-up and acceleration ramp-down. 
The Modified Acceleration Pulse (MAP) algorithm (see Algorithm 2) determines the control points for the speed ramp.
A speed cruise phase will be calculated if needed, with its duration set by \eqref{cruise time}.
\begin{equation}
\small
\begin{aligned}
\label{cruise time}
dt_\text{cruise}=dt-dt_\text{accpulse}
\end{aligned}
\end{equation}
\vspace{-0.6cm}

\begin{table}[h!]
  \centering
  \renewcommand{\arraystretch}{0.5} % Reset row height
  \colorbox{white!10}{\fbox{%
    \begin{tabular}{p{0.95\linewidth}} % 1 column, full width
      \multicolumn{1}{c}{\textbf{Algorithm 2.} Modified Acceleration Pulse (MAP)} \\
      \\
      \textbf{Input}: $v_1, v_2, a_{max}, j_{max},$ and $dt$.
      \\
      \\
      \textbf{Output}: Position, Velocity, Acceleration, and Jerk Profiles.
      \\
        \begin{enumerate}
        \item If $\left(d V<d V_{\min }\right.$ and $d t<d t_{\min }$ )
        
        \hspace{0.2cm} Set $dt_{\text{accpulse}}=dt$, and calculate $a_{\text {peak }}$ from \eqref{apeak}
        \begin{enumerate}
        \item If $\left(d V<a_{\text {peak }} d t\right)$ 
        
        Recalculate $dt$ from \eqref{apeak time}, and $a_{\text {peak}}$ from \eqref{apeak}
        \end{enumerate}
        
        Elseif $\left(d t<d t_{\min }\right)$
        
        \hspace{0.2cm} Set $dt_{\text{accpulse}}=dt$ and calculate $a_{\text {peak }}$ from \eqref{apeak}
        
        Elseif $\left(d V<d V_{\min }\right)$
        
        \hspace{0.2cm} Calculate $dt$ from \eqref{apeak time}, and $a_{\text {peak}}$ from \eqref{apeak}.
        \item Specify the time, position, velocity, acceleration, and jerk associated with the four quintic control points that constitute the segment of motion with nonzero acceleration.
        \vspace{-0.3cm}
        \end{enumerate}

    \end{tabular}%
  }}
\end{table}

\subsubsection{Generating Quintic Control Points}

The aforementioned methods are combined to generate a seamless trajectory between two waypoints, customized to the end conditions at each point. Each waypoint has a defined position and velocity, which must be achieved with zero acceleration and jerk. Given the specified time, initial and final velocities, and the joint's kinematic characteristics, the number of quintic segments will be determined using Algorithm 3.

\subsubsection{Trajectory Generation}

Employing the provided algorithms to compute the quintic control points, we then determine the coefficients of the corresponding polynomial for each segment. As per \eqref{polynomial}, six coefficients are calculated by taking the first and second derivatives to establish equations for velocity and acceleration, yielding six equations for the initial and final points. However, since absolute position can be measured from the environment, we reduce the equations to five by combining the position equations at the initial and final times to obtain the corresponding displacement. Setting the initial time to zero and the final time to \(dt\), we can solve \eqref{eqmat} to obtain the coefficient values.
\begin{equation}
\small
\begin{aligned}
\label{eqmat}
\begin{bmatrix}
    dt & dt^2 & dt^3 & dt^4 & dt^5 \\
    1 & 0 & 0 & 0 & 0 \\
    0 & 2 & 0 & 0 & 0 \\
    1 & 2dt & 3dt^2 & 4dt^3 & 5dt^4 \\
    0 & 2 & 6dt & 12dt^2 & 20dt^3
\end{bmatrix}
\begin{bmatrix}
    b_1 \\
    b_2 \\
    b_3 \\
    b_4 \\
    b_5
\end{bmatrix}
=
\begin{bmatrix}
    d \\
    v_1 \\
    a_1 \\
    v_2 \\
    a_2
\end{bmatrix}
\end{aligned}
\end{equation}
where $b_1,…,b_5$ are quintic coefficients, $d$ is the displacement, $v_1,v_2$ are the initial and final velocities, and $a_1,a_2$ are the initial and final acceleration respectively. The coefficients can be calculated easily by solving the equation.

\subsubsection{Safety Calculations}
As discussed earlier, we have used action masking to generate kinematically safe actions within our DRL framework. The output of the DRL network is a scalar $\bar{v}_t \in [-1,1]$ which must be linearly mapped to the interval $[v_{(t+1)_{\text{min}}}, v_{(t+1)_{\text{max}}}]$. Here, $v_{(t+1)_{\text{min}}}$ and $v_{(t+1)_{\text{max}}}$ respectively represent the minimum and maximum allowable velocities in that state, ensuring kinematically safe motion. This mapping is achieved using \eqref{mapping}.
\begin{equation}
\small
\begin{aligned}
\label{mapping}
v_{(t+1)} = v_{(t+1)_{\text{min}}} + \left(\frac{1 + \bar{v}_t}{2}\right) \times \left( v_{(t+1)_{\text{max}}} - v_{(t+1)_{\text{min}}} \right)
\end{aligned}
\end{equation}

\begin{figure}[h] % Adjust the value as needed
    \centering
     \hspace{-0.5cm}\includegraphics[width=0.5\textwidth, height=3.8cm]{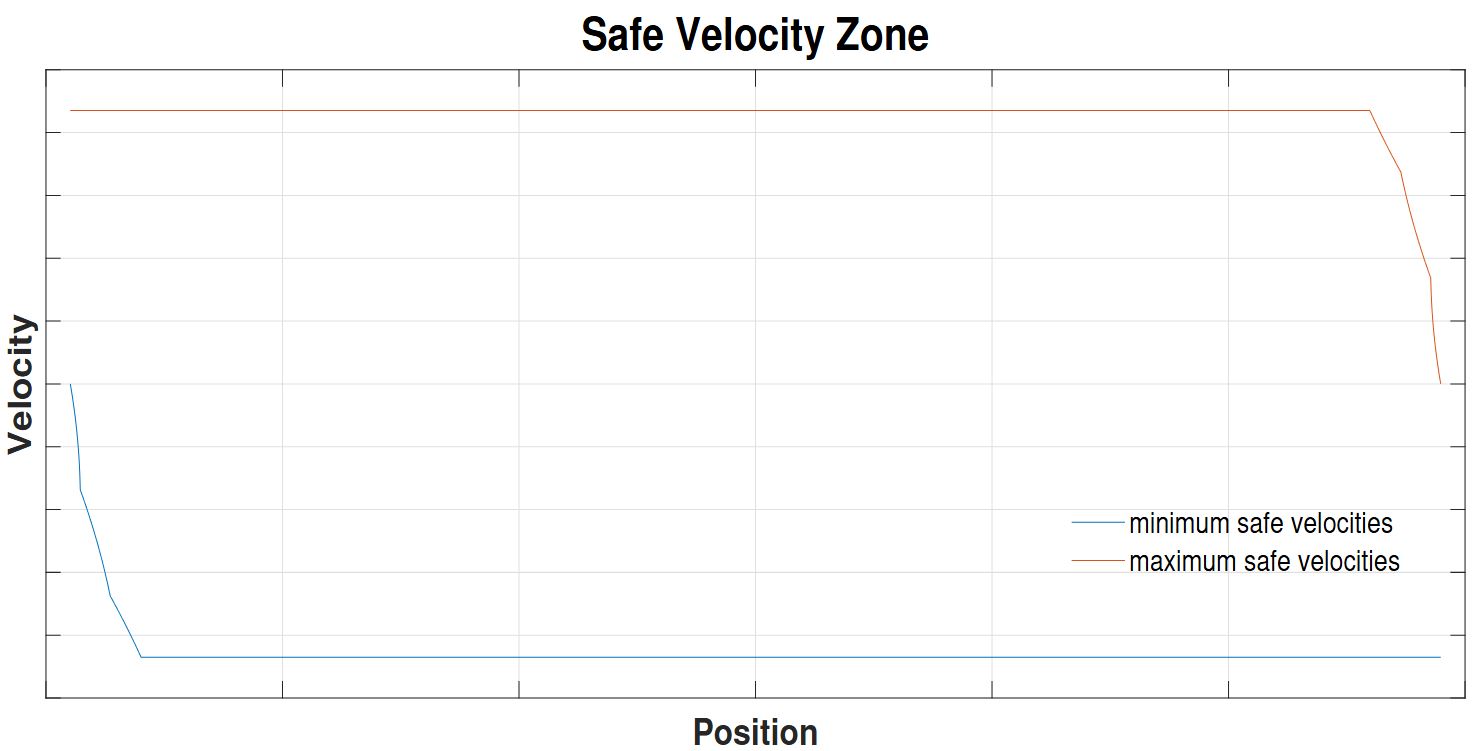}
    \caption{Safe velocity zone for a single joint.}
    \label{safezone}
\end{figure}

Kinematically safe behavior ensures joint motion meets the conditions in \eqref{constraint1} using a jerk-bounded velocity profile between DRL actions. As discussed in Section \ref{safety}, the JBTG calculates the achievable velocity range $[v1_{(t+1)_{\text{min}}}, v1_{(t+1)_{\text{max}}}]$ for the next step, ensuring compliance with velocity, acceleration, and jerk constraints.
To meet position constraints, we precompute safe velocity zones at various distances from the joint limits. By calculating braking distances for each velocity and setting the final velocity to zero, we generate pairs of braking distances and their corresponding maximum allowable velocities. This results in a safe velocity range of $[v2_{(t+1)_{\text{min}}},v2_{(t+1)_{\text{max}}}]$ at each point, as shown in Fig. \ref{safezone}.

However, to utilize this safe velocity zone, we need to know the joint's position. To achieve this, we first input the previously calculated values of $v1_{(t+1)_{\text{min}}}$ and $v1_{(t+1)_{\text{max}}}$ into the JBTG to estimate the joint's position in the next step. Using these estimated positions, we then calculate $v2_{(t+1)_{\text{max}}}$ based on the maximum safe velocity for the position corresponding to $v1_{(t+1)_{\text{max}}}$ and $v2_{(t+1)_{\text{min}}}$ based on the minimum safe velocity for the position estimated using $v1_{(t+1)_{\text{min}}}$. The final safe velocity range would be calculated as follows:
\begin{equation}
\small
\begin{aligned}
\label{3}
& [v_{(t+1)_{\text{min}}},v_{(t+1)_{\text{max}}}]= \\
& [max(v1_{(t+1)_{\text{min}}},v2_{(t+1)_{\text{min}}}), min(v1_{(t+1)_{\text{max}}},v2_{(t+1)_{\text{max}}})]
\end{aligned}
\end{equation}

\begin{table}[h!]
  \centering
  \renewcommand{\arraystretch}{0.5} % Adjust row height
  \colorbox{white!10}{\fbox{%
    \begin{tabular}{p{0.95\linewidth}} 
      \multicolumn{1}{c}{\textbf{Algorithm 3.} Calculation of quintic control points between two waypoints} \\
      \\
      \textbf{Input}: $v_1, v_2, v_{max}, a_{max}, j_{max},$ and $dt$.
      \\
      \\
      \textbf{Output}: quintic control points \vspace{0.1cm}
      \\
      We assume $v_2 > v_1$; if $v_2 < v_1$, we treat it as an increasing speed problem and make adjustments after calculating the trajectory.
      \\
      \\
        \textbf{Case 1:} If $v_2 \neq v_1$, $v_{\max} = v_2$.
          \begin{enumerate}
            \item If $dt > 2 \times dt_{\max}$ and $v_2 - v_1 > dV_{\min}$: \hspace{0.2cm}
            Use MSAP.
            \item Else: \hspace{0.2cm}
            Use MAP.
          \end{enumerate}

        \textbf{Case 2:} If $v_2 \neq v_1$, $v_{\max} > v_2$
          \begin{enumerate}
            \item If the required time to ramp up from $v_1$ to $v_2$ is less than $dt$, set $v_{\max} = v_2$. Go to Case 1.
            \item If $v_{\max} - v_2 > dV_{\min}$: \hspace{0.2cm}
            A minimum time, \( dt_{\text{limit}}\), is calculated as the time required to use MSAP to ramp from \(v_1\) to \(v_{\text{limit}}\) and then from \(v_{\text{limit}}\) to \(v_2\), where \(v_{\text{limit}}\) is the speed achieved by an acceleration pulse from \(v_2\) at \(a_{\max}\). Finally, \(dt_{\text{limit}}\) is compared to \(dt\) to decide if the trajectory will use MAPs or MSAPs.

            \begin{enumerate}
                \item If $dt > dt_{\text{limit}}$: \hspace{0.15cm}
                Use MSAPs to create speed ramps from $v_1$ to $v_{\text{peak}}$ and then from $v_{\text{peak}}$ to $v_2$, where $v_{\text{peak}}$ is the maximum speed achieved between the waypoints.

                \begin{enumerate}
                    \item If $dt$ is sufficiently large and is larger than the required time to reach $v_{\max}$: \hspace{0.1cm} set $v_{\text{peak}} = v_{\max}$ and a speed cruise is necessary at $v_{\max}$.

                    \item Else:\hspace{0.1cm}                  
                    $v_{\text{peak}} < v_{\max}$ should be calculated based on $dt$.

                \end{enumerate}

                \item Else if $dt < dt_{\text{limit}}$: \hspace{0.15cm}
                Utilizing at least one MAP, we ensure that \(v_{\text{peak}} < v_{\max}\). The value \(v_{\text{peak}}\) is obtained from a speed ramp from \(v_1\) to \(v_2\), followed by a pulse from \(v_2\) to \(v_{\text{peak}}\) and back. This value is then employed to generate the trajectory, which consists of a speed ramp from \(v_1\) to \(v_{\text{peak}}\) and another ramp from \(v_{\text{peak}}\) to \(v_2\).

            \end{enumerate}
            \item If $v_{\max} - v_2 < dV_{\min}$: \hspace{0.2cm}
            In this case, a MAP is required for the ramp down from $v_{\text{peak}}$ to $v_2$. Similar to Case 2.2, a limiting time, $dt_{\text{limit}}$, is calculated based on $v_{\text{peak}} = v_{\max}$.

            \begin{enumerate}
                \item If $dt$ is sufficiently large: \hspace{0.1cm}
                Set \(v_{\text{peak}} = v_{\max}\). A speed ramp brings \(v_1\) to \(v_{\max}\), followed by a speed cruise at \(v_{\text{peak}}\).

                \item Else: \hspace{0.1cm}
                Case 2.2b is used.

            \end{enumerate}

          \end{enumerate}

        \textbf{Case 3:} $v_2 = v_1$, $v_{\max} = v_2$.

          A speed cruise at $v_{\max}$. 

        \textbf{Case 4:} $v_2 = v_1$, $v_{\max} > v_2$.

        For this case, we do the same as in Case 2, with the difference that the ramp-up and ramp-down profiles are symmetrical.
        \\
        \\

      {\scriptsize * We require \(v_2\) to be nonnegative. If both \(v_1\) and \(v_2\) are nonpositive, we take their absolute values and recheck the condition. If \(v_1 > 0\) and \(v_2 < 0\), we negate the velocities and recheck the initial condition, adjusting the trajectory as needed. If the final velocity is nonnegative, the profile will move toward \(v_{\max}\) for maximum positive displacement; otherwise, it will adjust toward \(v_{\min}\) for maximum negative displacement.}
      \vspace{-0.3cm}

    \end{tabular}%
  }}
\end{table}

\subsection{Robust Control Design Framework}
We now define a subsystem-based tracking transformation as follows:
\begin{equation}
\small
\begin{aligned}
\label{304}
& z_1(t)=e_1(t), \hspace{0.2cm} z_2(t)=e_2(t)-u_v
\end{aligned}
\end{equation}
where $u_v$ is a virtual control to adjust position and velocity tracking and can be defined as in \cite{shahna2024integrating}:
\begin{equation}
\small
\begin{aligned}
\label{305}
u_v (t)=-\frac{1}{2}\left(k_1+k_2 \hat{\phi}_1(t)\right) z_1(t)
\end{aligned}
\end{equation}
where $k_1$, $k_2$ are positive constants and $\hat{\phi}_1$ is the adaptive law which will be defined in \eqref{307}. Therefore, based on \eqref{303}, \eqref{304}, and \eqref{305}, we obtain:
\begin{equation}
\small
\begin{aligned}
\label{306}
\dot{z}_1 (t) =& z_2(t)+f_1 + d_1 + u_v\\
\dot{z}_2(t) =& A \hspace{0.03cm}u(t)+f_2+ d_2  - \ddot{x}_r (t) - \dot{u}_v
\end{aligned}
\end{equation}
Now, we define the adaptive law for the position
\begin{equation}
\small
\begin{aligned}
\label{307}
\frac{d \hat{\phi}_1(t)}{d t}=-k_3 k_4 \hat{\phi}_1(t)+\frac{1}{2} k_2 k_3\left|z_1(t)\right|^2
\end{aligned}
\end{equation}
where $k_3$ and $k_4$ are positive constants. The error of adaptive law can be defined as:
\begin{equation}
\small
\begin{aligned}
\label{308}
\tilde{\phi}_1=\hat{\phi}_1-\phi_1^*
\end{aligned}
\end{equation}
From \eqref{307} and \eqref{308}, we obtain:
\begin{equation}
\small
\begin{aligned}
\label{309}
\frac{d \tilde{\phi}_1(t)}{d t}=-k_3 k_4 \tilde{\phi}_1(t)+\frac{1}{2} k_2 k_3\left|z_1(t)\right|^2-k_3 k_4 {\phi}_1^*
\end{aligned}
\end{equation}
Now, we define the control input signal as:
\begin{equation}
\small
\begin{aligned}
\label{310}
u(t) & =-z_{1}(t)-\frac{1}{2}\left(k_5+k_6 \hat{\phi}_2(t)\right) z_2(t)
\end{aligned}
\end{equation}
where $k_5$ and $k_6$ are positive constants and $\hat{\phi}_2$ is the second adaptive parameter. Similarly, we can have the second adaptive law as:
\begin{equation}
\small
\begin{aligned}
\label{311}
\frac{d \hat{\phi}_2(t)}{d t}=-k_7 k_8 \hat{\phi}_2(t)+\frac{1}{2} k_6 k_7\left|z_2(t)\right|^2
\end{aligned}
\end{equation}
where $k_7$ and $k_8$ are positive constants. Similarly, we can define the error of adaptive law:
\begin{equation}
\small
\begin{aligned}
\label{312}
\tilde{\phi}_2=\hat{\phi}_2-\phi_2^*
\end{aligned}
\end{equation}
From \eqref{312} and \eqref{311}:
\begin{equation}
\small
\begin{aligned}
\label{313}
\frac{d \tilde{\phi}_2(t)}{d t}=-k_7 k_8 \tilde{\phi}_1(t)+\frac{1}{2} k_6 k_7\left|z_1(t)\right|^2-k_7 k_8 {\phi}_1^*
\end{aligned}
\end{equation}
A Lyapunov function is suggested as follows:
\begin{equation}
\small
\begin{aligned}
\label{2013}
&V_{1} =\frac{1}{2} \hspace{0.1cm} [{z}^2_1 + k_3^{-1} \tilde{\phi}^2_1]
\end{aligned}
\end{equation} 
After differentiating $V_1$, we obtain:
\begin{equation}
\small
\begin{aligned}
\label{equation: 21}
\dot{V}_{1}=   z_1 [z_2(t)+f_1 + d_1 + u_v] + k_3^{-1} \dot{\tilde{\phi}}_1 {\tilde{\phi}}_1 
\end{aligned}
\end{equation} 
By considering the description of \eqref{300} and \eqref{304}, we obtain:
\begin{equation}
\small
\begin{aligned}
\label{eq3401}
\dot{V}_{1}=&   z_1 z_2(t)+z_1 f_1 + z_1 d_1 -\frac{1}{2}\left(k_1+k_2 \hat{\phi}_1(t)\right) z^2_1\\
&+ k_3^{-1} \dot{\tilde{\phi}}_1 {\tilde{\phi}}_1 
\end{aligned}
\end{equation} 
Assume $\xi_1$ and ${d}_1^* \in \mathbb{R}^+$ are unknown positive constants assigning the bound of uncertainties and disturbances. We define a bounded and continuous function $M_1: \mathbb{R} \rightarrow \mathbb{R}^+$ generating positive values as:
\begin{equation}
\small
\begin{aligned}
\label{eq3501}
&\|f_1\| \hspace{0.1cm} \leq \xi_1 \hspace{0.1cm}M_1\hspace{0.1cm}, \hspace{0.2cm} \|{d_1}\| \hspace{0.1cm} \leq  {d}_1^*
\end{aligned}
\end{equation}
Then, from \eqref{eq3401} and \eqref{eq3501}:
\begin{equation}
\small
\begin{aligned}
\label{equation: 21}
\dot{V}_{1} \leq&  z_1 z_2(t)+|z_1| \xi_1 \hspace{0.1cm}M_1 + |z_1| {d}_1^* -\frac{1}{2}k_1 z^2_1-\frac{1}{2}k_2 \hat{\phi}_1 z^2_1\\
&+ k_3^{-1} \dot{\tilde{\phi}}_1 {\tilde{\phi}}_1
\end{aligned}
\end{equation}
By mathematical manipulation, we can obtain:
\begin{equation}
\small
\begin{aligned}
\label{equation: 21}
\dot{V}_{1} \leq&  z_1 z_2(t)+ \frac{1}{4} \beta_1^{-1} \hspace{0.1cm}M_1^2+ \beta_1 |z_1|^2 \xi^2_1 \hspace{0.1cm} + \beta_2 (|z_1| {d}_1^*)^2 \\
&+ \frac{1}{4} \beta^{-1}_2 -\frac{1}{2}k_1 z^2_1-\frac{1}{2}k_2 \hat{\phi}_1 z^2_1+ k_3^{-1} \dot{\tilde{\phi}}_1 {\tilde{\phi}}_1
\end{aligned}
\end{equation} 
By inserting \eqref{309}, we have:
\begin{equation}
\small
\begin{aligned}
\label{equation: 21}
\dot{V}_{1} \leq&  z_1 z_2(t)+ \frac{1}{4} \beta_1^{-1} \hspace{0.1cm}M_1^2+ \beta_1 |z_1|^2 \xi^2_1 \hspace{0.1cm} + \beta_2 (|z_1| {d}_1^*)^2 \\
&+ \frac{1}{4} \beta^{-1}_2 -\frac{1}{2}k_1 z^2_1-\frac{1}{2}k_2 \hat{\phi}_1 z^2_1- k_4 \tilde{\phi}_1^2\\
&+\frac{1}{2} k_2 {\tilde{\phi}}_1 \left|z_1(t)\right|^2- k_4 {\phi}_1^* {\tilde{\phi}}_1
\end{aligned}
\end{equation} 
The unknown constant of the adaption law is defined:
\begin{equation}
\small
\begin{aligned}
\label{eq40}
\phi_1^*=& \frac{2}{k_2} ({\beta_2} {d_1^*}^2  + \beta_1 \xi_1^2)
\end{aligned}
\end{equation}
where
\begin{equation}
\small
\begin{aligned}
\label{equation: 21}
\dot{V}_{1} \leq&  z_1 z_2(t)+ \frac{1}{4} \beta_1^{-1} \hspace{0.1cm}M_1^2+ \frac{1}{4} \beta^{-1}_2 -\frac{1}{2}k_1 z^2_1- k_4 \tilde{\phi}_1^2\\
&- k_4 {\phi}_1^* {\tilde{\phi}}_1
\end{aligned}
\end{equation} 
By mathematical manipulation, we obtain:
\begin{equation}
\small
\begin{aligned}
\label{equation: 21}
\dot{V}_{1} \leq&  z_1 z_2(t)+ \frac{1}{4} \beta_1^{-1} \hspace{0.1cm}M_1^2 + \frac{1}{4} \beta^{-1}_2 -\frac{1}{2}k_1 z^2_1- \frac{1}{2}k_4 \tilde{\phi}_1^2\\
&+ \frac{1}{2} k_4 {{\phi}_1^*}^2
\end{aligned}
\end{equation} 
considering \eqref{2013}:
\begin{equation}
\small
\begin{aligned}
\label{equation: 2005}
\dot{V_{1}}\leq& z_1 z_2(t) -\psi_1 V_1+ \frac{1}{4} \beta_1^{-1} \hspace{0.1cm}M_1^2 + \frac{1}{4} \beta^{-1}_2+ \frac{1}{2} k_4 {{\phi}_1^*}^2
\end{aligned}
\end{equation} 
where
\begin{equation}
\small
\begin{aligned}
\label{eq68}
\psi_{1}=\min \left[\begin{array}{lll}
k_1, & k_3 k_4
\end{array}\right]
\end{aligned}
\end{equation}
Similarly, we can define:
\begin{equation}
\small
\begin{aligned}
\label{equation: 20}
&V_{2} =V_1 + \frac{1}{2} \hspace{0.1cm} [\frac{1}{A}{z}^2_2 + k_7^{-1} \tilde{\phi}^2_2]
\end{aligned}
\end{equation} 
and after differentiating $V_2$, we obtain:
\begin{equation}
\small
\begin{aligned}
\label{equation: 21}
\dot{V}_{2}=   \dot{V}_1 + \frac{1}{A} z_2 [A u(t)+f_2 + d_2 -\ddot{x}_r - \dot{u}_v] + k_7^{-1} \dot{\tilde{\phi}}_2 {\tilde{\phi}}_2 
\end{aligned}
\end{equation} 
By considering the description of \eqref{300} and \eqref{304}, we obtain:
\begin{equation}
\small
\begin{aligned}
\label{equation: 21}
\dot{V}_{2}=&   \dot{V}_1 + z_2 u(t)+\frac{1}{A}z_2 f_2 + \frac{1}{A}z_2 d_2 -\frac{1}{A}z_2 \ddot{x}_r - \frac{1}{A} z_2 \dot{u}_v\\
&+ k_7^{-1} \dot{\tilde{\phi}}_2 {\tilde{\phi}}_2 
\end{aligned}
\end{equation} 
Assume $ f_2^* = \frac{1}{A}f_2 -\frac{1}{A}\dot{u}_v$ and $\bar{d}_2 = \frac{1}{A}d_2 - \frac{1}{A} \ddot{x}_r$. We can have:
\begin{equation}
\small
\begin{aligned}
\label{equation: 21}
\dot{V}_{2}\leq&   z_1 z_2(t) -\psi_1 V_1+ \frac{1}{4} \beta_1^{-1} \hspace{0.1cm}M_1^2 + \frac{1}{4} \beta^{-1}_2+ \frac{1}{2} k_4 {{\phi}_1^*}^2 \\
&-  z_1 z_2-\frac{1}{2} k_5 z_2^2-\frac{1}{2} k_6 z_2^2 \hat{\phi}_2+z_2 f_2^* + z_2 \bar{d}_2\\
&+ k_7^{-1} \dot{\tilde{\phi}}_2 {\tilde{\phi}}_2 
\end{aligned}
\end{equation} 
In addition, assume $d_2^*$ and $ \xi_2 \in \mathbb{R}^+$ are unknown positive constants assigning the bound of uncertainties and disturbances. We define a bounded and continuous function $M_2: \mathbb{R} \rightarrow \mathbb{R}^+$ generating positive values as:
\begin{equation}
\small
\begin{aligned}
\label{eq35}
&\|f_2^*\| \hspace{0.1cm} \leq \xi_2 \hspace{0.1cm}M_2\hspace{0.1cm}, \hspace{0.2cm} \|{\bar{d}_2}\| \hspace{0.1cm} \leq  {d}_2^*
\end{aligned}
\end{equation}
Similar to \eqref{equation: 2005}, we can obtain:
\begin{equation}
\small
\begin{aligned}
\label{equation: 21}
\dot{V_{2}}\leq& -\psi_2 V_2+ \frac{1}{4} \beta_3^{-1} \hspace{0.1cm}M_2^2 + \frac{1}{4} \beta_1^{-1} \hspace{0.1cm}M_1^2 \\
&+ \frac{1}{4} \beta^{-1}_2+ \frac{1}{4} \beta^{-1}_4+ \frac{1}{2} k_4 {{\phi}_1^*}^2+ \frac{1}{2} k_8 {{\phi}_2^*}^2
\end{aligned}
\end{equation}
where the positive constant of the adaptive law is selected as:
\begin{equation}
\small
\begin{aligned}
\label{eq40}
\phi_2^*=& \frac{2}{k_4} ({\beta_4} {d_2^*}^2  + \beta_3 \xi_2^2), \hspace{0.2cm}{\psi_{2}=\min \left[\begin{array}{lll}
k_5, & k_7 k_8
\end{array}\right]}
\end{aligned}
\end{equation}
Thus, by defining the system's Lyapunov function as $V = V_1 + V_2$, following \cite{shahna2024integrating} and \cite{heydari2024robust}, the system with the proposed control is uniformly exponentially stable.

\section{SIMULATIONS AND RESULTS}

\begin{figure}[h] % Adjust the value as needed
    \centering
    \includegraphics[width=0.3\textwidth, height=3.3cm]{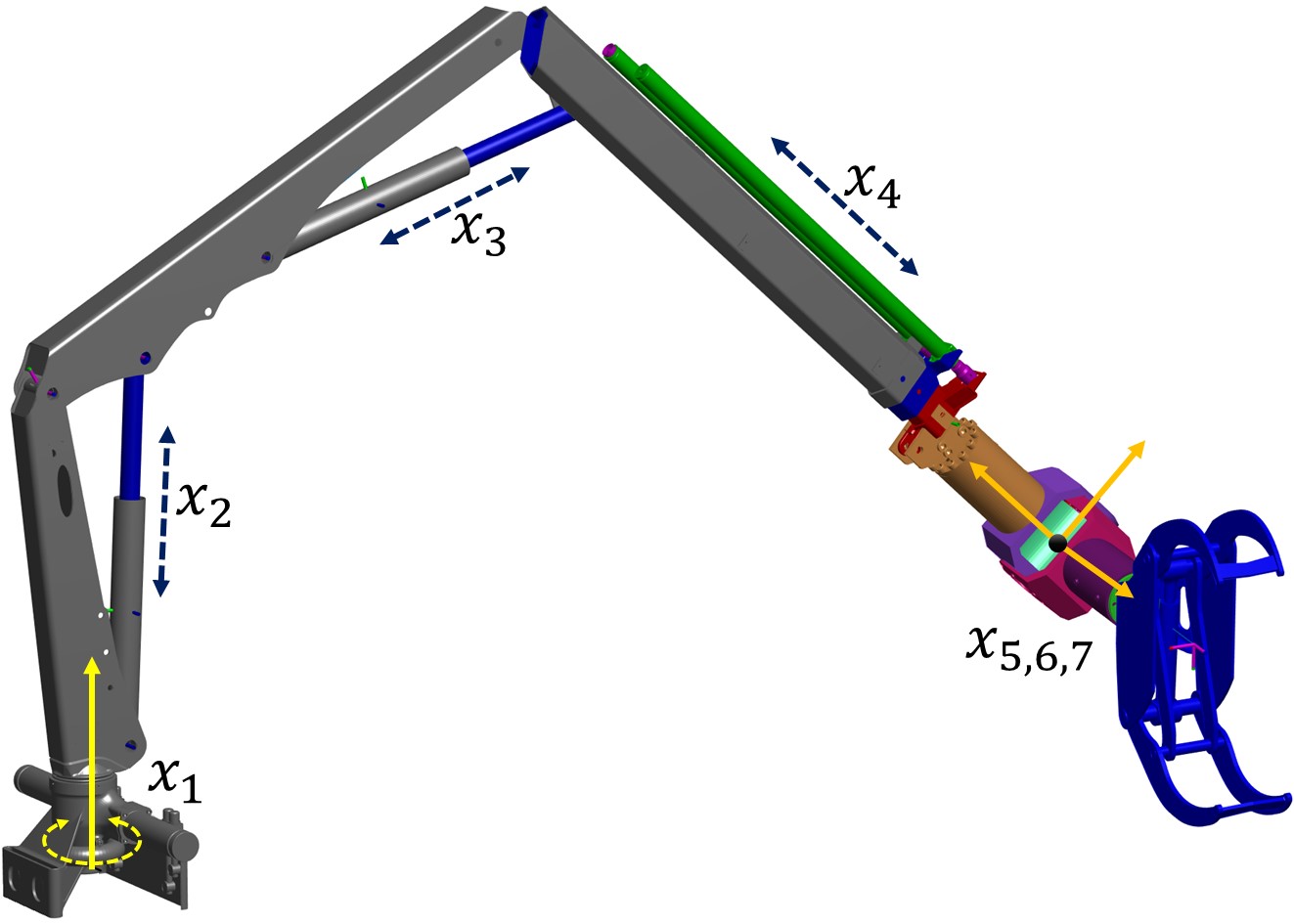}
    \caption{Heavy-duty robot model \cite{alvaro2024analytical}}
    \label{hiab}
\end{figure}

For the evaluation, we applied our proposed method to a heavy-duty industrial robot model \cite{alvaro2024analytical} (Fig. \ref{hiab}) performing a reaching task, integrating the modified JBTG within the SAC algorithm as the high-level controller for online trajectory planning, alongside a robust subsystem-based adaptive controller as the low-level controller. The heavy-duty robot model was specifically chosen to demonstrate the effectiveness of our method, given its sensitivity to nonsmooth trajectories due to its lower natural frequencies.

The DRL agent, detailed in Table \ref{DRLagent}, operates at $20\ Hz$, receiving model states every $0.05$ seconds and generating corresponding actions. 
However, the JBTG operates at a frequency of $1\ kHz$, and the low-level controller at $2\ kHz$.

\begin{table}[h]
  %\captionsetup{position=top}
  \caption{DRL Agent Parameters}
  \label{DRLagent}
  \centering
  \renewcommand{\arraystretch}{0.8}
  \scriptsize
  \begin{tabular}{lclc}
    \toprule
    {\textbf{Parameters}} & {\textbf{Value}} & {\textbf{Parameters}} & {\textbf{Value}} \\
    \midrule
    Batch size & \bm{$128$} & Target smoothing factor & \bm{$0.001$} \\
    Experience buffer length & \bm{$1000000$} & Initial random steps & \bm{$10000$} \\
    Discount factor & \bm{$0.99$} & Training episodes & \bm{$3000$} \\
    Time step & \bm{$0.05$} & Max time steps/episode & \bm{$300$} \\
    Learning rate & \bm{$0.0001$} \\
    \bottomrule
  \end{tabular}
\end{table}

The parameters related to the JBTG are defined as follows: $x \in [0.14, 0.50] \ m$, $v_{max}=0.15\ m/s$, $a_{max}=1\ m/s^2$, and $j_{max}=100\ m/s^3$. The time interval for the JBTG corresponds to the DRL agent's step time of \(0.05 \, \text{s}\).

To operate the manipulator within its workspace, we focus on the three prismatic joints ($x_2, x_3, x_4$ in Fig. \ref{hiab}) out of a total of seven, ensuring that the end-effector movements remain in the same plane. The simulation features varied initial and target positions, allowing the agent to effectively learn to navigate between different configurations by the end of training. We define reaching accuracy as $5 \ \text{cm}$ at a tip velocity of less than $0.1 \ \text{m/s}$. When the manipulator reaches this threshold of the target, the episode terminates, and the agent receives the reaching reward.

\begin{figure}[h] % Adjust the value as needed
    \centering
     \hspace{-0.5cm}\includegraphics[width=0.36\textwidth, height=3.5cm]{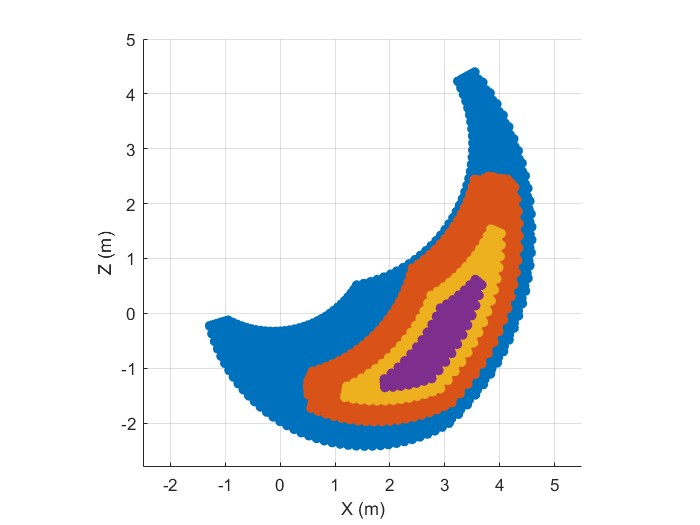}
    \caption{Workspace of the manipulator tip with designated regions of CL}
    \label{workspace}
\end{figure}

To implement curriculum learning in our DRL agent, we divide the workspace into multiple regions. In Fig. \ref{workspace}, the blue region represents the manipulator's entire workspace, while the orange region is the target area for final operations. Initially, we set the end-effector's positions within the purple region, which is easier to reach due to its shorter distance. Once the robot masters this area, we expand the training to the yellow region and then to the orange one. Proficiency in the orange region signifies that we have reached our goal. Importantly, each region encompasses the previous ones, ensuring the agent retains its learned policies.

The low-level control system manages all seven DoFs. In addition to the three prismatic joints directed by the DRL, the remaining four joints are necessary to maintain the initial position according to the desired trajectories. This control system must also compensate for gravity and external disturbances affecting these joints during movement.

\begin{figure}[h] % Adjust the value as needed
    \centering
    \hspace{-0.3cm}\includegraphics[width=0.42\textwidth, height=3.1cm]{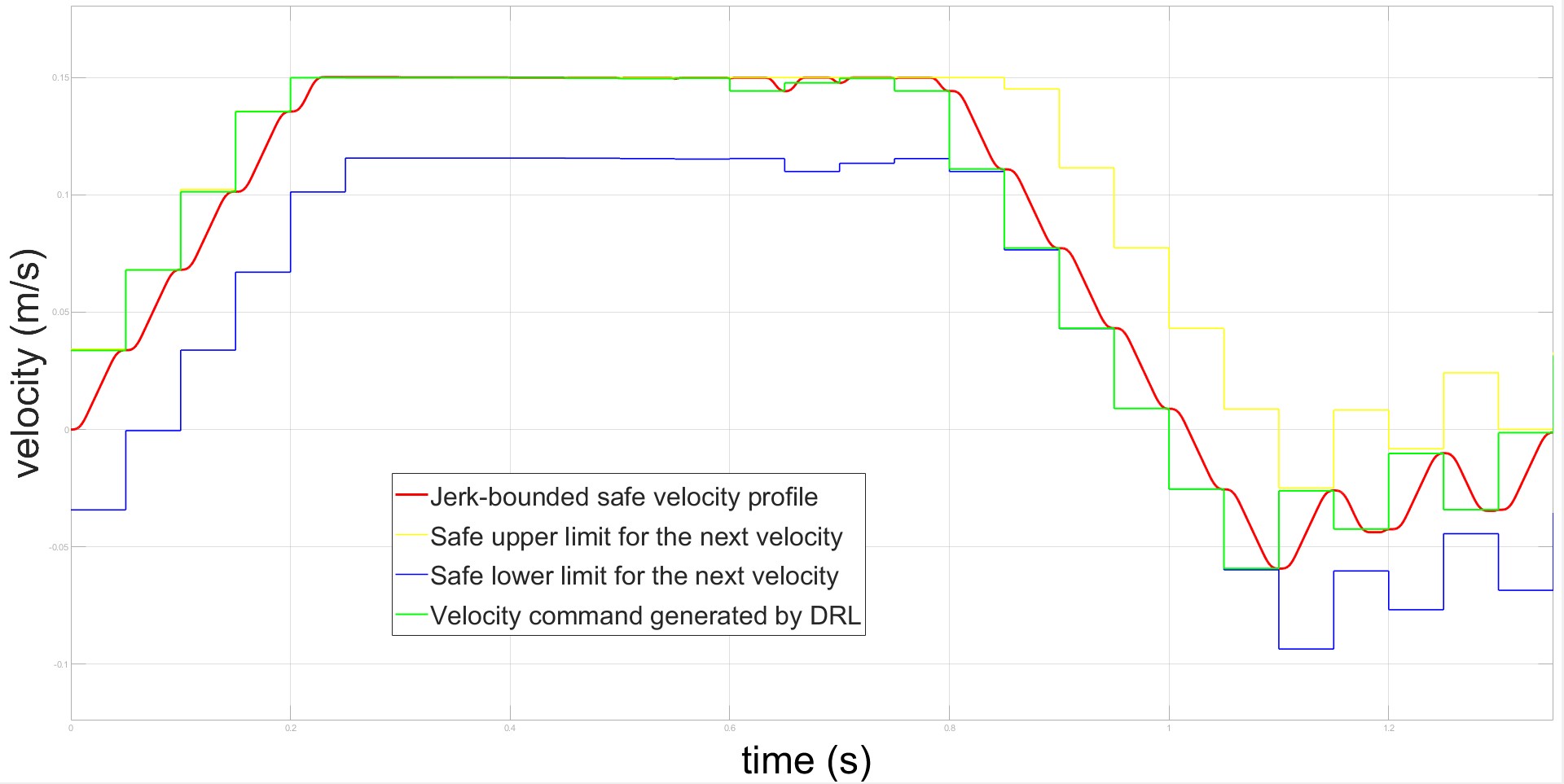}
    \caption{Generated velocity profile along with safe velocity ranges}
    \label{velocity profile}
\end{figure}

The graphs in Fig. \ref{velocity profile} depict the generated trajectory and the velocity range for a single joint during a reaching task after training is complete. At the beginning of each DRL time step, the maximum (yellow), minimum (blue), and mapped selected velocity (green) are set and remain fixed throughout the time step. The red curve represents the safe jerk-bounded velocity profile, computed based on the initial and final velocities for that DRL time step. This trajectory is passed to the low-level controller for tracking.

\begin{figure}[h] % Adjust the value as needed
    \centering
     \hspace{-0.5cm}\includegraphics[width=0.46\textwidth, height=3.6cm]{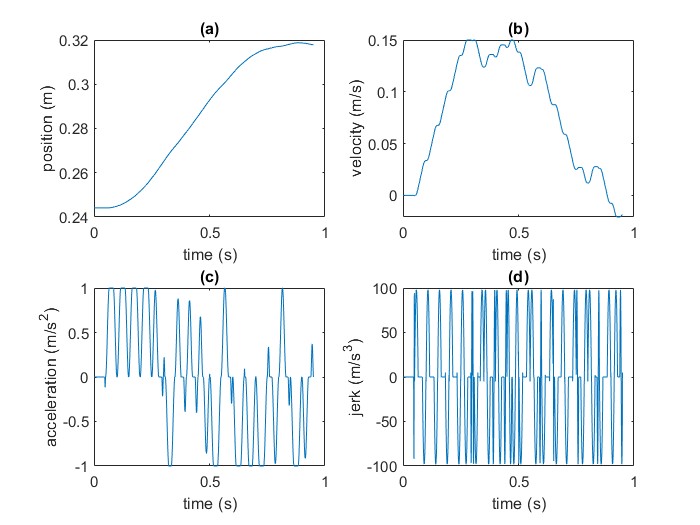}
    \caption{Graphs of data collected from a successful reaching task for one joint: (a) Position, (b) Velocity, (c) Acceleration, and (d) Jerk}
    \label{reaching profile}
\end{figure}

Fig. \ref{reaching profile} shows the position, velocity, acceleration, and jerk profiles of a single joint during a successful attempt in which the manipulator reached the target position.

\begin{figure}[h] % Adjust the value as needed
    \centering
     \hspace{-0.4cm}\includegraphics[width=0.46\textwidth, height=3.6cm]{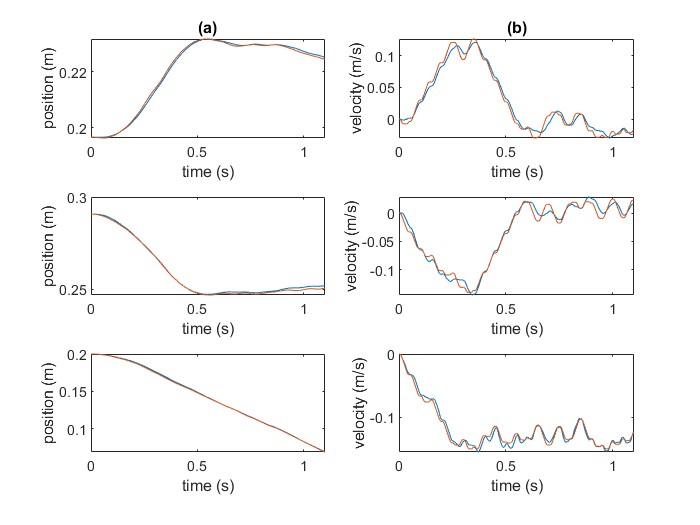}
    \caption{Performance of low-level control to track the jerk-bounded (a) position and (b) velocity profiles}
    \label{tracking}
\end{figure}

Fig. \ref{tracking} demonstrates the effectiveness of the robust low-level controller in tracking the jerk-bounded velocity and position profiles generated by the DRL in three joints $x_2$, $x_3$, and $x_4$. The red trajectories represent the reference jerk-bounded trajectory profiles, while the blue trajectories show the actual motion profiles. The remaining joints, through the implementation of the robust low-level control, maintained the initial positions, with an error of less than $3$ mm.

\section{CONCLUSIONS}

This paper presents a novel framework that integrates a JBTG with a robust low-level controller within a DRL framework, significantly enhancing the safety, stability, and efficiency of robotic manipulators. By incorporating jerk-bounded profiles, we ensure smooth and continuous motion, mitigating the risk of abrupt actions that could compromise safety. The integration of pre-calculated safe velocity zones for joint positions further enhances the system's reliability by preventing unsafe trajectories and adhering to kinematic constraints.
Simulation results confirm that this approach effectively balances safety and stable performance, resulting in smoother trajectories, with the active participation of the robust low-level control strategy in the learning procedure.
Future work will aim to enhance this approach for more complex scenarios involving self-collisions and external obstacles. Moreover, efforts will focus on ensuring the system's kinematic and dynamic safety, facilitating practical implementation without safety violations.

\bibliographystyle{IEEEtran}
\bibliography{main.bib}

\end{document}